\documentclass[a4paper,fleqn,usenatbib]{mnras}
\usepackage{amssymb,amsmath,graphicx,psfrag,mathtools}
\usepackage[T1]{fontenc} 
\usepackage{ae,aecompl} 
\usepackage{url}
\title[FRB as Products of Accretion Disc Funnels]{FRB as Products of
Accretion Disc Funnels}
\author[J. I. Katz]{
J. I. Katz,$^{1}$\thanks{E-mail katz@wuphys.wustl.edu} 
\\
$^{1}$Department of Physics and McDonnell Center for the Space Sciences,
Washington University, St. Louis, Mo. 63130 USA 
}
\date{Accepted XXX.  Received YYY; in original form ZZZ} 
\pubyear{2017} 
\date{\today}
\begin{document} 
\label{firstpage} 
\pagerange{\pageref{firstpage}--\pageref{lastpage}} 
\maketitle 
\begin{abstract}
The repeating FRB 121102, the only FRB with an accurately determined
position, is associated with a variable persistent radio source consistent
with a low luminosity active galactic nucleus.  I suggest that FRB originate
in the accretion disc funnels of intermediate mass black holes.  Narrowly
collimated radiation is emitted along the wandering instantaneous angular
momentum axis of accreted matter.  We observe this emission as a fast radio
burst when it sweeps across the direction to the observer.  This model
constrains the mass of the black hole to values below those of galactic
nuclei.  It predicts, in contrast to neutron star (pulsar or SGR) models,
that repeating FRB will not be periodic and will be co-located with
persistent but variable radio sources resulting from the off-axis emission.
The model is analogous, on smaller spatial, lower mass and accretion rate
and shorter temporal scales, to AGN making double radio sources, with FRB
corresponding to blazars in which the jets point toward us.  
\end{abstract}
\section{Introduction}
Most FRB models have been based on neutron stars.  FRB have been suggested
to be higher energy versions of pulsar pulses, powered by neutron star
rotational energy, or the result of SGR-like dissipation of magnetostatic
energy (see \citet{K16a} for a review).  Pulsar models require extreme
values of rotation rate and magnetic field to account for FRB peak powers if
these are bounded by their spin-down power (see, however, \citet{K17b}).  A
possible loophole, a wandering collimated beam \citep{K17a}, mitigates this
problem but narrow collimation may be inconsistent with a diverging fan of
magnetic field lines.  The open field lines of a rapidly rotating neutron
star diverge widely even near its surface.  SGR-based models must be
reconciled with the non-detection of a FRB during the giant flare of SGR
1820-06 \citep{TKP16,K16b}.

This paper proposes an entirely different model based on accretion onto an
``intermediate mass'' ($\sim 10^2\text{--}10^6 M_\odot$) black hole.  This
model scales down an AGN rather than scaling up (in energy) a radio pulsar.
The model still involves a wandering narrow beam to mitigate the power
requirement of isotropic emission.  The comparatively large black hole mass
permits a high accretion rate and accretional luminosity, although these
cannot be estimated without more detailed knowledge of the environment.  The
narrow funnels of a thick accretion disc \citep{FKR02} confine the
relativistic \citep{K14} radiating particles to a narrow jet, and their
radiation to a narrow beam.

The assumed narrowly collimated beam is analogous to the jets emitted by
AGN, micro-quasars, GRB, SS433 and even protostars.  The beam wanders
\citep{K17a} because it is aligned with the disc's instantaneous angular
momentum axis that follows the angular momentum of accreting matter.
Wandering disc angular momentum is plausible for a intermediate-mass black
hole accreting from a chaotic medium such as a giant molecular cloud in a
starburst galaxy.  This is unlike a supermassive black hole in a galactic
nucleus that accretes from a disc aligned with the galactic symmetry plane.
When the beam sweeps across the direction to an observer, a burst is
observed.  The requirement that intrinsic burst durations not exceed a
millisecond constrains the mass of the black hole.
\section{Implications of the Persistent Source}
The repeating FRB 121102 is associated with a variable persistent radio
source \citep{C17} and identified with a dwarf galaxy at $z = 0.193$
\citep{T17}.  The persistent radio source ``varies by tens of percent on
days timescales'' \citep{C17}; they measured rates of change as high as
7\%/day, implying a characteristic time of 14 days, and suggest it is a low
luminosity AGN.  The persistent source has a projected separation from the
FRB of $< 40$ pc \citep{M17} in a dwarf galaxy about 100 times larger, 
indicating co-location in space and a causal connection rather than an
accidental association.

Rapid variability argues against pulsar wind nebula and supernova remnant
models \citep{MKM16,KM17,M17,MBM17,DWY17,B17,W17} of the persistent source.
Radio emission from the best studied young Galactic pulsar wind nebula, the
Crab Nebula, \citep{AR85} and young supernova remnant, Cas A, \citep{DAO74}
decays very gradually, at a rate $\alpha/T$, where $T$ is their age and
$\alpha$ is a factor ${\cal O}(1)$.  If the decay is a power law function of
time  $\alpha$ is the exponent.  There is no evidence of shorter term
variability in these objects, nor would that be expected because of the
large size of the nebul\ae\ and remnants and the long synchrotron energy
loss time of electrons that radiate at radio frequencies.  The variability
of the persistent source, consisting of a flare above a steadier background,
is reminiscent of the behavor of AGN rather than of scintillation, but even
if variability is the result of scintillation an AGN remains consistent with
the observations.

Even a much younger hypothetical pulsar wind nebula or supernova remnant
associated with FRB 121102, whose age is at least four years and size is
$\gtrsim 10^{17}$ cm (at typical supernova expansion speed $\sim 10^9$
cm/s), would not vary nearly as fast as the persistent counterpart.  The
time scale of variation could be shorter if the bulk
radiating plasma were expanding towards us relativistically, but this is
possible only for a pulsar wind nebula unconfined by surrounding matter.
Its radius would be $\ge 4$ light-years and free expansion would be
inconsistent with any plausible ambient density; there is no evidence of
behavior of this kind in Galactic objects.

Rapid variability and the observed \citep{C17} nonthermal radio spectrum
suggest an accreting black hole, as in an AGN.  The source's comparatively
low luminosity $\nu L_\nu \approx 6 \times 10^{38}$ ergs/s at $\nu = 3$ GHz,
as well as its presence in a dwarf galaxy of mass estimated to be only
4--7$\times 10^7\,M_\odot$, indicate a much lower black hole mass than in
AGN.  These values may be compared to the luminosity at 2.6 GHz of 3C273
$\nu L_\nu = 7 \times 10^{43}$ ergs/s \citep{S08} and black hole mass of
$\sim 7 \times 10^9 M_\odot$ \citep{PT05} suggesting a giant elliptical host
galaxy of mass $\sim 10^{12}\text{--}10^{13}M_\odot$ \citep{M03}.  Rough
scaling by a factor $\sim 10^5$ suggests the origin of FRB 121102 in
accretion onto an intermediate mass black hole.  
\section{Making FRB} 
Accretion, whether on stellar-mass objects or supermassive black holes, is
observed to produce fluctuating but not impulsive radiated power.  Hence I
adopt the assumption of \citet{K17a} that the emission is very narrowly
collimated and that the observed bursts result from the sweeping of this
narrow beam across the direction to the observer.  This assumption was made
in order to mitigate the demands on the energetics of a pulsar-like source,
but here is applied to a source powered by accretion onto a black hole.

The most accurately measured jets, those of SS433, wander in angle by a few
degrees about their mean precessional motion \citep{KP82}.  Because SS433
accretes from its companion in a binary star whose orbital plane is fixed,
this indicates that at least some wandering is an intrinsic property of
accretion discs.  If FRB accrete from dense interstellar or protostellar
clouds that are likely to be ``turbulent'' or ``chaotic'' in density and
velocity in regions of active star formation, the direction of the accreted
angular momentum may wander through large angles.
\section{Energetics}
In this model the radiated energy is supplied by accretion.  The mean power
of bursts depends on their duty factor $D$, which is only known (and even
there only very roughly) for the repeating FRB 121102, for which it is
$\sim 10^{36}$ ergs/s during active periods when bursts are separated by
intervals $\sim 1$ minute.  This is small compared to the power $\nu L_\nu
\sim 6 \times 10^{38}$ ergs/s at 3 GHz of the persistent source.  If the
persistent power is derived from accretion onto a black hole, the accretion
rate must be $\gtrsim 10^{19}$ g/s.  The inequality results from the
omission of unknown luminosity outside the 3 GHz band, the unknown
efficiency (\citet{W17} suggested it may be substantial) of producing radio
radiation and from the fact that in super-Eddington flows the accretion rate
is unbounded while the emergent luminosity saturates around the Eddington
limit as a result of radiation trapping \citep{K77}.

A black hole of mass $M$ moving at supersonic speed $v$ through a
homogeneous cloud of density $\rho$ at rest accretes at the rate
\begin{equation}
\begin{split}
{\dot M} &\approx \rho {(GM)^2 \over v^3} \\ &\approx 10^{19} \left({M
\over 100 M_\odot}\right)^2 \left({n_H \over 10^5\,\text{cm}^{-3}}\right)
\left({10\,\text{km/s} \over v}\right)^3\ \text{g/s},
\end{split}
\end{equation}
where $n_H$ is the density of hydrogen atoms.  The second and third factors
in parentheses are plausibly ${\cal O}(1)$ for a dense molecular cloud.  The
first factor in parentheses may be $\gg 1$.  This classical result (due to
Bondi) is unlikely to be applicable because interstellar gas is surely
spatially heterogeneous and may have significant velocity gradients; if it
were applicable, then the accreted matter would have no angular momentum
about the black hole and would not form a disc (or jet or beam).

The fact that accreting black holes, either of stellar mass or supermassive
in AGN, produce jets and often appear to be surrounded by discs, establishes
the importance of heterogeneity of the gas distribution and of angular
momentum to the accretion process.  Unfortunately, this is not understood in
detail and no predictive theory exists; for example, the low luminosity of
Sgr A$^*$ at the center of our Galaxy is unexplained.  We have almost no
information about the environment of FRB 121102 beyond its presence in a
dwarf galaxy (for example, is it in a dense molecular cloud?), so it is not
possible to say more than that accretion sufficient to power the persistent
source is not excluded.
\section{Opening Angles}
The duty factor $D$ of a repeating FRB (for FRB that have not been observed
to repeat only an upper bound on $D$, approximately the length of the burst
divided by the duration of observation, can be set) is related to the
opening angle $\theta$:
\begin{equation}
\theta \approx \sqrt{\Omega D},
\end{equation}
where $\Omega$ is the solid angle over which the beam wanders.
Optimistically taking $\Omega = 4 \pi$ yields $\theta \approx 10^{-2}$ for
the repeating FRB 121102, and $\theta \lesssim 3 \times 10^{-4}$ for the
best observed other FRB.

These estimates may be compared to observed jet opening angles \citep{H91}.
In the microquasar/X-ray source Sco X-1 an upper bound of about 1$^\circ$
may be estimated from the size of the radio lobes (the jet itself is not
detected) \citep{F83}, while in other microquasars jets may be as narrow as
$\sim 2^\circ$ \citep{MJ06}.  Parsec-scale jets in AGN may have opening
angles of $\sim 1^\circ$ \citep{BP84,O89,B94,P99,P09}.  

These values are larger (much larger in the case of non-repeating FRB) than
the beam widths required to explain the observed duty factors.  However, the
observed opening angles are those of radiation produced by the interaction of
the accelerated jet with surrounding matter.  This interaction region may be
a turbulent layer much wider than the jet itself, broadened as turbulence
propagates into the ambient medium.  It is likely to be wider than the core
of energetic particles or radiation that is not directly observed.  Even in
SS433, where the observed emission lines are ascribed to the entire
(subrelativistic) jet, the radiation may be emitted by ions outside a much
narrower jet core that are repeatedly ionized and recombined by interaction
with surrounding matter that scatters them from the core.
\section{Bounds on Black Hole Masses}
The characteristic time scale or duration of a burst is the time required
for a beam of width $\theta$ to sweep through its width
\begin{equation}
\label{tchar}
\Delta t \gtrsim {2GM \theta \over c^3 \alpha} \sim {2GM \over c^3 \alpha
\gamma},
\end{equation}
where $\gamma \sim 1/\theta$ is the Lorentz factor of the accelerated
particles and $\alpha$ is the friction factor (ratio of viscous stress to
isotropic pressure) of the disc \citep{FKR02}.  Here we have multiplied the
characteristic inner disc relaxation time $2GM/(c^3 \alpha)$ by $\theta$ to
allow for the fact that reorientation by an angle $\theta$, if the disc is
supplied by a turbulent flow, is sufficient to begin or end a burst.

Estimating, as before, $\theta \sim \sqrt{4 \pi D}$, Eq.~\ref{tchar} becomes
\begin{equation}
{M \over M_\odot} \lesssim 100 {\Delta t_{-3} \alpha \over \sqrt{4 \pi D}},
\end{equation}
where $\Delta t_{-3} \equiv \Delta t/10^{-3}\,\text{s}$.  Typical estimates
of $\alpha$ are $\sim 10^{-2}$ \citep{FKR02}, but measurements of the rapid
``nodding'' motion of the jets of SS433 indicate an effective $\alpha \sim
1$ for the propagation of orientational information to the jets
\citep{KAMG82}.

Substituting $D \sim 10^{-5}$ for the repeating FRB 121102 and a possible
$D \sim 10^{-8}$ for a non-repeating FRB yields bounds on the masses of
their black holes $M$
\begin{equation}
{M \over M_\odot} \lesssim
\begin{cases}
10^4 \Delta t_{-3} \alpha & \text{FRB 121102}\\
3 \times 10^5 \Delta t_{-3} \alpha & \text{non-repeaters}.
\end{cases}
\end{equation}
The bounds are less strict for smaller values of $D$.  By definition, only
an upper limit on $D$ can be set for a non-repeater, so that if $D$ is much
smaller than the nominal $10^{-8}$ the bound on $M$ for non-repeaters is
relaxed.

The bound for FRB 121102 is firmer, although its numerical value depends
(not sensitively) on the uncertain evaluation of $D$.  Despite this
uncertainty, we conclude that in the accretion funnel model the black hole
has a mass intermediate between the stellar masses of black holes in binary
X-ray sources and the supermassive black holes powering AGN at the centers
of galaxies.
\section{Discussion}
The black hole accretion funnel model predicts:
\begin{enumerate}
\item FRB are accompanied by persistent but variable radio sources, their
out-of-beam radiation.  These persistent sources are analogous to the
radiation of AGN whose jets are not pointed toward us (AGN that are not
blazars).  An example of such an AGN with its jets and radio lobes
fortuitously appears near the bottom of Fig.~2a of \citet{C17}.
\item Repeating FRB will not be periodic, because (unlike in models based on
rotating neutron stars) there is no clock.
\item FRB will be associated with very soft X-ray/extreme UV sources
produced by accretion discs around intermediate mass black holes.  Such
radiation is strongly absorbed in the Galactic plane and difficult to
observe from FRB 121102 at low Galactic latitude, but may be observable
from other FRB.
\end{enumerate}

No model of FRB, or even of classic radio pulsars, has yet satisfactorily
explained their coherent emission, so this must be left to future work.
\section*{Acknowledgements}
I thank J. Eilek for useful discussions.

\bsp 
\label{lastpage} 
\end{document}